\documentstyle[epsfig]{EuroPhys}
\input EuroMacr
\begin{document}

\euro{**}{*}{?-?}{**}
\Date{}
\shorttitle{M. RAIBLE et al. : AMORPHOUS THIN FILM GROWTH}

\title{Amorphous thin film growth: theory compared with experiment}
\author{M. Raible, S. G. Mayr*, S. J. Linz,
M. Moske**, P. H\"anggi, K. Samwer*}
\institute{
Institut f\" ur Physik, Universit\" at Augsburg,
D-86135 Augsburg, Germany\newline
*present address: I. Physikalisches Institut, Universit\"at G\"ottingen,
Bunsenstr. 9,\\D-37073 G\"ottingen, Germany\newline
**present address: Research center caesar,
Friedensplatz 16, D-53111 Bonn, Germany}
\rec{**}{**}

\pacs{
\Pacs{68}{35Bs}{Surface structure and topography}
\Pacs{61}{43Dq}{Amorphous alloys}
      }

\maketitle

\begin{abstract}
Experimental results on amorphous ZrAlCu thin film growth and the 
dynamics of the surface morphology as predicted from a minimal nonlinear 
stochastic deposition equation are analysed and compared. Key points 
of this study are (i) an estimation procedure for coefficients entering 
into the growth equation and (ii) a detailed analysis and interpretation of the time 
evolution of the correlation length and the surface roughness. The results 
corroborate the usefulness of the deposition equation as a tool for 
studying amorphous growth processes.
\end{abstract}
\renewcommand{\thefootnote}{(\arabic{footnote})}

%\pacs{PACS numbers: 05.30.-d, 05.40.+j, 33.80.Be}
%%%%%%%%%%%%%%%%%%%%%%%% INTRODUCTION %%%%%%%%%%%%%%%%%%%%%%
\vspace{1.5ex}{\em Introduction.} --
During the last decade the study of the kinetics of surface growth 
processes has attracted  considerable interest (cf. the reviews \cite{Bar95}).
The dynamics of the surface morphology e.g. in amorphous thin film growth
is dominated by the interplay of roughening, smoothening, and pattern forming
processes. On the microscopic level, these processes are governed by the highly
complex and only partly understood interaction of the depositing particles with
the already condensed surface atoms. Despite the complexity of the growth
processes on the atomic scale, experiments on the slightly coarser mesoscopic
scale typically reveal some sort of regularity of the surface morphology with
some superimposed small-scale stochastics \cite{Rei97,Sal95}. This, in turn, 
indicates that the machinery of coarse-grained continuum models based on 
phenomenologically motivated stochastic growth equations \cite{Bar95} is
a useful tool for the understanding and interpretation of the growth dynamics.
In particular, amorphous thin film growth represents an attractive  testing 
ground for the validation of such phenomenological models; 
this is mainly due to the spatially isotropic nature of the amorphous
structure at this scale and the lack of long range ordering.\\
Our objective is a detailed comparison between a
stochastic nonlinear evolution equation for amorphous
thin film growth \cite{wir99} and experimental results 
on the surface morphology of ZrAlCu films prepared by physical vapor deposition
that are analysed using
scanning  tunneling microscopy (STM). Using the aforementioned theoretical
approach,  we develop a method to 
estimate the phenomenological parameters that is based on the
evolution for short times.
We also investigate whether the finite apex angle of the cone-like STM tip
used in the experiment does affect the comparison between theory and experimental
findings. In particular, we demonstrate  that the results for both the surface
roughness $w$ (i.e. the root mean square deviation of the relative height
fluctuations) and correlation length $R_c$ (i.e. the typical length scale over
which height fluctuations are correlated) of our full nonlinear growth equation
yield very good agreement with the presented experimental data, with only minor
modifications due to the finite apex angle.
This clearly corroborates the usefulness of our modeling approach.

\begin{figure}[htb]
\begin{center}
\epsfxsize 6.5cm
\epsfbox{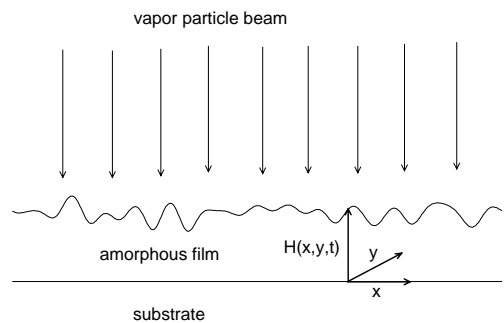}
\end{center}
\caption{Sketch of a physical vapor deposition experiment
at normal incidence for amorphous film growth on a substrate.}
\label{fig:DEPOS}
\end{figure}

\vspace{1.5ex}{\em Experiments.} --
The glassy ZrAlCu films (composition 
Zr$_{65}$Al$_{7.5}$Cu$_{27.5}$) are prepared in ultra high vacuum by physical
vapor deposition on oxidized Si wafers using a total deposition rate of 
$F=0.79nm/s$ (electron beam evaporation of the pure elements, each source
independently rate controlled). Due to the geometrical arrangements, the
particle flux is almost  perpendicular to the substrate which is rotated during
deposition (for a sketch of the experiment setup cf. fig.~\ref{fig:DEPOS}). The
surface profiles of the films (sample area ($200\times200$)nm$^2$) are
analysed  {\em in situ} by scanning tunneling microscopy using a tungsten 
tip with an apex angle of ca. $40^{\circ}$ (for further
details cf. Ref. \cite{Rei97}). The film composition is verified by Auger
electron spectroscopy. Additionally performed X-ray diffraction and
differential scanning calorimetry (DSC)
measurements ensure the amorphicity of the investigated films. From the
STM data, the correlation length $R_c$ and the surface roughness $w$ (for their
definition, see below) are determined for various layer thicknesses up to
$480$nm. The experimental results correspond to the diamond symbols
depicted in fig.~\ref{fig:bildrcw1} and ~\ref{fig:bildrcw2} below.
At large layer thicknesses ($\ge 100nm$)
each such symbol represents a measurement with a different sample.

\vspace{1.5ex}{\em Theoretical modeling.} --
To accomplish a theoretical description of the time-evolution of an
initially flat
surface $H(\vec{x},t)$, where $H$ denotes the $z$-coordinate of the growing
surface at the position $\vec{x}=(x,y)$ and time $t$ (cf. also
fig.~\ref{fig:DEPOS}), we take advantage of
the well-established phenomenological approach that is based on
stochastic nonlinear partial differential equations \cite{Bar95}, i.e.
\begin{equation}
\partial_tH=G(H,\nabla H)+F+\eta.
\label{spd}
\end{equation}
In eq. (\ref{spd}), $G$ denotes a functional of the surface
height and its local derivatives. The detailed functional form of $G$ depends
crucially on the considered experimental setup and the kinetics of the
deposition process.
$F$ denotes the mean deposition rate and $\eta(\vec{x},t)$ is the
related deposition noise that determines the fluctuations of the deposition
process around its mean $F$. These fluctuations are assumed to be Gaussian
white,
\begin{equation}
\left<\eta(\vec{x},t)\right>=0;\quad
\left<\eta(\vec{x},t)\eta(\vec{y},t')\right>=
2D\delta^2(\vec{x}-\vec{y})\delta(t-t')
\label{noise}
\end{equation}
where the brackets denote ensemble averaging. It proofs useful to
introduce the height profile $h(\vec{x},t)=H(\vec{x},t)-Ft$.
If the deposition process has any excess velocity, there is a nonlinear relation
between the mean growth or layer thickness $\left<H\right>$ and time, i.e.
$\left<H\right>(t)=Ft+\left<h\right>(t)$ with $\left<h\right>(t)\neq 0$.\\
The simplest nonlinear amorphous thin film growth equation of the
functional form (\ref{spd}) that incorporates (i) the physical symmetries
such as rotation and reflection invariance in the plane perpendicular to the
growth direction, cf. fig.~\ref{fig:DEPOS},
(ii) no particle desorption, and (iii) the potentiality of local density variations
in the amorphously grown material yields in terms of a
low-order expansion of $G$ in the gradients of the surface profile
$h(\vec{x},t)$ the result
\begin{equation}
\partial_th=a_1\nabla^2h+a_2\nabla^4h
+a_3\nabla^2(\nabla h)^2+a_4(\nabla h)^2
+a_5\det\left(\begin{array}{cc}
\partial_x^2h&\partial_y\partial_xh\\
\partial_x\partial_yh&\partial_y^2h
\end{array}\right)+\eta.
\label{sto1}
\end{equation}
with $a_i(i=1,\ldots,5)$ being scalar material-dependent coefficients. In eq.
(\ref{sto1}), the first and the fifth term on the r.h.s. are directly related
to the deflection of the initially perpendicular incident particles due to the
interatomic forces between the surface atoms and the incident
particles. For an indication of the relevance of this effect we
refer to the recent
experimental study in \cite{Dij99}. The coefficients $a_1$ and $a_5$
are determined by the relations $a_1=-Fb$ and $a_5=Fb^2$, where $b$ is the
difference between the
typical range of the interatomic forces and the equilibrium distance of the
adatoms to the surface. Because $b$ is small for
the aforementioned experimental setup (typically of the order $10^{-1}nm$)
the term proportional to $a_5$ in eq. (\ref{sto1}) can safely be neglected.
The second and the third term on the r.h.s. of eq. (\ref{sto1}) are
related to the known microscopic mechanisms of (i) the surface diffusion
suggested by Mullins \cite{Mul57}
and (ii) the equilibration of the inhomogeneous concentration of the
diffusing particles on the surface, as suggested in \cite{Vil91,Mos97}.
Moreover, the coefficients $a_2$ and $a_3$ are negative.
The coefficient $a_3$ reads $a_3=-Fl^2/8$, where $l^2$
is the mean square of the diffusion length of the particles. 
This characteristic dependence for amorphous growth is similar in nature to a
term appearing for crystalline growth \cite{Kal97}; it differs,
however, due to the absence
of an additional length scale, the typical height of crystalline
terraces. The fourth term on the r.h.s. of eq. (\ref{sto1}) is of
Kardar-Parisi-Zhang (KPZ) form \cite{KPZ86}. It is due to the potential dependence
of the local density on the surface slope $\nabla h$, i.e.
$[\rho(\nabla h)]^{-1}=\rho_0^{-1}[1+(a_4/F)(\nabla h)^2+{\cal O}((\nabla h)^4)]$,
with $a_4$ being necessarily positive because of the additional volume increase
caused by oblique particle incidence. Note also that a finite $a_4$ results
in a finite excess velocity.\\
The experimentally detected correlation length $R_c(t)$ and surface roughness
$w(t)$ are determined by the height-height correlation function
\begin{equation}
C(r,t)=
\left.\left<\frac{1}{L^2}\int d^2x
\left(h(\vec{x},t)-\overline{h}\right)
\left(h(\vec{x}+\vec{r},t)-\overline{h}\right)\right>
\right|_{|\vec{r}|=r}
\label{corr}
\end{equation}
where $\overline{h}$ is the spatial average of the surface profile
$\overline{h}=\frac{1}{L^2}\int d^2yh(\vec{y},t)$ and $L^2$ the sample area.
Specifically, $R_c(t)$ is
given by the first maximum of $C(r,t)$ occuring at nonzero $r$ and the square
of the surface roughness results by taking the
limit $r=0$ in $C(r,t)$, i.e. $w^2(t)=C(0,t)$. To integrate
numerically the growth equation, a forward-backward finite difference method on a
quadratic lattice combined with an Euler algorithm in time and periodic
boundary conditions on a quadratic area $[0,L]^2$ have been invoked \cite{Mos91}.

\vspace{1.5ex}{\em Linear model equation.} --
In order to make contact with the aforementioned experiment, we first
extrapolate the values of the parameters $a_1$, $a_2$, and $D$. The initial
stages of the deposition process
(corresponding to layer thicknesses $\left<H\right>\le 240nm$ in the experiment),
which started from a
flat substrate, are dominated by the temporal evolution of the linear limit of
eq. (\ref{sto1}),
\begin{equation} \partial_th=a_1\nabla^2h+a_2\nabla^4h+\eta.
\label{lin1}
\end{equation}
The growth of the Fourier modes of the surface profile is determined by
$\partial_t\tilde{h}(\vec{k},t)
=\sigma(k)\tilde{h}(\vec{k},t)+\tilde{\eta}(\vec{k},t)$ where
$\sigma(k)=-a_1k^2+a_2k^4$ denotes the growth rate of the Fourier modes.
Therefore, the wave number $k_c$ belonging to the maximum of the
growth coefficient $\sigma(k)$ reads $k_c=\sqrt{a_1/2a_2}$.
The correlation length $R_c(t)$ of the surface profile
$h(\vec{x},t)$ arising from the linearized equation (\ref{lin1}) follows
initially a $t^{1/4}$-law, whereas it saturates at later stages into
$R_c(t)=7.0156/k_c=7.0156\sqrt{2a_2/a_1}$ when the
critical mode with the wave number $k_c$ dominates. Since a saturation of
$R_c(t)$ with increasing layer thickness is also observed in the experiment
for layer thicknesses larger than $200nm$ (cf. the diamond symbols in
fig.~\ref{fig:bildrcw1}a), the ratio $a_2/a_1$ can be roughly estimated.
The surface roughness $w(t)$ at later stages follows
an $\exp[\sigma(k_c)t]=\exp(-a_1^2t/4a_2)$-behaviour when the critical
mode dominates. By a comparison with the experimentally observed increase of
$w(t)$ (cf. the diamond symbols in fig.~\ref{fig:bildrcw1}b)
during the time interval
when the layer thickness $\left<H\right>$ is between $30nm$ and $240nm$ the ratio
$a_1^2/a_2$ can be estimated. From the ratios $a_2/a_1$ and $a_1^2/a_2$
the coefficients $a_1$ and $a_2$ can be deduced.
The height-height correlation function $C(r,t)$
that arises from the linearized equation (\ref{lin1}) is determined by
%\begin{equation}
%C(r,t)=\left.\frac{D}{L^2}\sum_{\vec{k}\neq 0}\exp(i\vec{k}\cdot\vec{r})
%\frac{\exp\left[2(-a_1k^2+a_2k^4)t\right]-1}{-a_1k^2+a_2k^4}\right|_{|\vec{r}|=r}.
%\end{equation}
%Since all possible wave vectors $\vec{k}$ have the form
%$\vec{k}=(2\pi n_x/L,2\pi n_y/L)$ where $n_x$ and $n_y$ are integer numbers,
%$C(r,t)$ converges in the limit of large $L$, yielding
\begin{equation}
C(r,t)=\left.\frac{D}{(2\pi)^2}\int d^2k\exp(i\vec{k}\cdot\vec{r})
\frac{\exp\left[2(-a_1k^2+a_2k^4)t\right]-1}{-a_1k^2+a_2k^4}\right|_{|\vec{r}|=r}.
\label{corr1}
\end{equation}
From $C(r,t)$ the surface roughness
$w(t)$ and the correlation length $R_c(t)$ can be determined.
This makes it possible to determine more precisely the
coefficients of eq. (\ref{lin1}). The physical dimensions of these
coefficients are $[a_1]=\overline{l}^2/\overline{t}$,
$[a_2]=\overline{l}^4/\overline{t}$,
and $[D]=\overline{h}^2\overline{l}^2/\overline{t}$ where
$\overline{l}$ denotes a length unit, $\overline{t}$ a time unit,
and $\overline{h}$ a height unit, respectively.
Therefore, $\sqrt{a_2/a_1}$ is a length constant, $|a_2/a_1^2|$ is a time
constant, and $\sqrt{D/|a_1|}$ is a height constant. Changing $\sqrt{a_2/a_1}$
by an arbitrary factor would change all lengths by the same factor. So we
calculated $w(t)$ and $R_c(t)$ by means of eq. (\ref{corr1}) with the
approximately determined coefficients $a_1$ and $a_2$ and an arbitrary $D$.
Then, we changed $R_c$, time $t$, and $w$ by independent factors and
$\sqrt{a_2/a_1}$, $|a_2/a_1^2|$, and $\sqrt{D/|a_1|}$ by the same factors,
respectively, until $R_c(t)$ and $w(t)$ were in accordance with the
experimental result. By means of $\sqrt{a_2/a_1}$, $|a_2/a_1^2|$, and
$\sqrt{D/|a_1|}$ we evaluated $a_1$, $a_2$, and $D$. The parameters that fit
the experimental results best, are found to read
\begin{equation}
a_1=-0.0826nm^2/s,\quad
a_2=-0.319nm^4/s,\quad
D=0.0174nm^4/s.
\label{para}
\end{equation}
The dotted lines in
fig.~\ref{fig:bildrcw1}a and ~\ref{fig:bildrcw1}b depict the corresponding
numerical simulations of eq. (\ref{lin1}). There is obviously good agreement
with the experimental results for layer thicknesses $\left<H\right>\le 240nm$.\\
Next, we investigate possible effects caused by the STM tip.
Although the detailed shape of the STM tip is known only roughly a reasonable
model, which refers to the worst case situation,
consists of a combination of a taper shank with an apex angle of ca.
$40^{\circ}$ and a top cone having a considerably wider apex angle of
approximately $\alpha=100^{\circ}$. Only the latter part is of relevance for
the detection of the surface profile. The effect of the STM tip is that it
maps the height profile $h(\vec{x},t)$ from eq. (\ref{lin1}) on the "scanned"
height profile $\hat{h}(\vec{x},t)$.
The solid line in fig.~\ref{fig:bildrcw1}b depicts
the corresponding finding for this worst case of a STM tip with an apex angle
of $100^{\circ}$. The major effect of the finite tip angle
results in a slow down of the growth of the
surface roughness $w(t)$. This is caused by the scanning procedure that
can not fully resolve the grooves of the height profile $h(\vec{x},t)$.
As can be deduced from fig.~\ref{fig:bildrcw1}b, however,
the linear growth eq. (\ref{lin1}) (even including the tip angle)
is not sufficient to describe the later stages of
amorphous thin film growth.

\vspace{1.5ex}{\em Nonlinear model equation.} --
We next investigate the physical relevance
of the nonlinear terms $a_3\nabla^2(\nabla h)^2$ and $a_4(\nabla h)^2$
on the growth dynamics, i.e. we analyse the nonlinear growth model 
\begin{equation}
\partial_th=a_1\nabla^2h+a_2\nabla^4h+a_3\nabla^2(\nabla h)^2+a_4(\nabla h)^2+\eta.
\label{sto2}
\end{equation}
The parameter choice that fits the experimental results best, is found to read
\begin{equation}
a_3=-0.10nm^3/s\quad\mbox{and}\quad a_4=0.055nm/s,
\end{equation}
where the parameters given in eq. (\ref{para}) were used.
The corresponding results for the correlation length $R_c(t)$ and the surface 
roughness $w(t)$ of eq. (\ref{sto2}) are depicted in fig.~\ref{fig:bildrcw2}
with (solid lines) and without (dotted lines) the inclusion of the effect of the
STM tip. We now obtain good agreement with the experimental data
also at larger layer thicknesses. Moreover, the finite apex angle of the tip
yields only a minor effect (up to the largest experimental layer thickness
$\approx 480nm$), see the barely visible dotted line in fig.~\ref{fig:bildrcw2}b.
As a general consequence, the two nonlinear terms result
in a drastic slow down of the increase of the surface
roughness  $w(t)$  with time or layer thickness. The larger the absolute values
of $a_3$ and $a_4$ are, the stronger is this slow down. Nevertheless, the
inclusion of {\em both} nonlinearities is necessary to obtain a consistent agreement
with the experimental data.\\
In particular, by setting $a_4=0$ we need to decrease further $a_3$ in order
to fit reasonably well the data. Nevertheless, the surface roughness $w(t)$ is
now either too large at large thickness $\left<H\right>=480nm$, or too
small at $\left<H\right>=360nm$ (not shown). The inclusion of the finite tip
geometry does not cure this finding. Omitting $a_3$ we find an overshoot for
the roughness $w(t)$ above $\left<H\right>\approx 240nm$.
In addition, we note that the correlation length $R_c(t)$ ceases to exist
above $\left<H\right>\approx 300nm$, because the first maximum of
the height-height-correlation $C(r,t)$ vanishes. These features are depicted
by the dashed lines in fig.~\ref{fig:bildrcw2}a and ~\ref{fig:bildrcw2}b.
Fig.~\ref{fig:bildrcw2}b also depicts that the experimental data for $w(t)$
are strongly scattered at largest layer thickness. Thus, a precise estimation
of the nonlinear coefficients is difficult. We estimate that the
coefficient $a_3$ ranges at most between $-0.13nm^3/s\le a_3\le -0.08nm^3/s$
and likewise $0.04nm/s\le a_4\le 0.07nm/s$.
Consistent with the fitting procedure in
fig.~\ref{fig:bildrcw1}a and ~\ref{fig:bildrcw1}b we find that the crossover
from linear to nonlinear growth behaviour sets in at
$\left<H\right>\approx 240nm$. An inclusion of the small correction
proportional to $a_5$, see eq. (\ref{sto1}), does not quantitatively impact
our results: For the largest layer thickness we find an improvement of
maximal $1.0\%$ for the roughness $w(t)$ and $2.5\%$ for the correlation
length $R_c(t)$.

\vspace{1.5ex}{\em Discussion.} --
The aforementioned extrapolation of the parameters $a_1$, $a_2$, $a_3$, $a_4$, and $D$
also allows for additional microscopic estimates.
(i) Since $a_1=-Fb$, the typical range $b$ of the interaction
between the surface atoms and the particles to be deposited, is about $0.1nm$,
i.e. it approximately equals the size of the radii ($0.2nm$) of the surface atoms.
(ii) Since $a_3=-Fl^2/8$, the diffusion length $l$ must be in the range
of $1.0nm$. This substantiates that the deposited particles experience a
surface diffusion on a nanometer scale and do not just stick at the
places where they hit the surface.
(iii) If the particles arrive independently
on the surface, the deposition noise is related to the particle volume $\Omega$
and the mean deposition rate $F$ by $2D=F\Omega$, yielding (by use of eq.
(\ref{para})) $\Omega=0.04nm^3$. This agrees within a factor of two with
the averaged particle volume of ZrAlCu.
(iv) The necessity of the inclusion of the term proportional to $a_4$ indicates
that the local density of the growing film varies with the surface slope: On
an inclined surface area the local density is decreased by
$\rho(\nabla h)=\rho_0/\gamma$ with $\gamma=1+(a_4/F)(\nabla h)^2$,
where $a_4/F$ is in the range of $0.07$.
Note that these predicted finite density variations are physically compatible with the
small diffusion length $l$ of two to three atom diameters.
At largest layer thickness ($\approx 480nm$) this local density reduction
(averaged over the surface) possesses a mean $\overline{\gamma}=1.021$
and a standard deviation
$\left(\overline{\left(\gamma-\overline{\gamma}\right)^2}\right)^{1/2}=0.017$.
The maximum of $\gamma$ on
the same surface (at $L=200nm$) is typically of the order of $1.12$.
At the same time, $\gamma$ averaged over the whole film equals
$\left<H\right>/(Ft)=1.01$.

\vspace{1.5ex}{\em Conclusions}.-- Starting from experimental data for
amorphous ZrAlCu thin film growth, the phenomenological growth equation in
(\ref{sto1}), and using a step-by-step parameter identification procedure, 
we have shown that a quantitative agreement between theory and experiment 
is achieved. Based on our
comparison between theory and experiment, we conclude that (i) eq.
(\ref{sto1})  constitutes a valid theoretical model for amorphous thin film
growth  (at least up to the considered layer thicknesses) and (ii) that an 
interpretation of the data necessarily requires the inclusion of the two
nonlinear terms $\nabla^2(\nabla h)^2$ and $(\nabla h)^2$
in the stochastic growth equation (\ref{sto1}). 
Several questions, however, remain open for future studies: (a) From the
available  experimental data it is not clear whether the correlation length
$R_c(t)$ and  the surface roughness $w(t)$ saturate for
larger layer thicknesses. (b) The validity of the proposed approach
needs to be tested more thoroughly by direct comparison with experimental data
for spatio-temporal quantities such as the height-height correlation function.
(c) A further experimental challenge presents the validation of the
theoretically predicted local density variations.

\vspace{1.5ex}{\em Acknowledgement}.-- This work has been supported by the
DFG-Sonderforschungsbereich 438 M\"unchen/Augsburg, TP A1. The authors are
indebted to B. Reinker for providing some additional experimental results.
The authors like
to thank Prof. D. E. Wolf for his most helpful and constructive remarks.

\begin{figure} \begin{center}
\hspace*{\fill}
\epsfxsize 7.0cm
\epsfbox{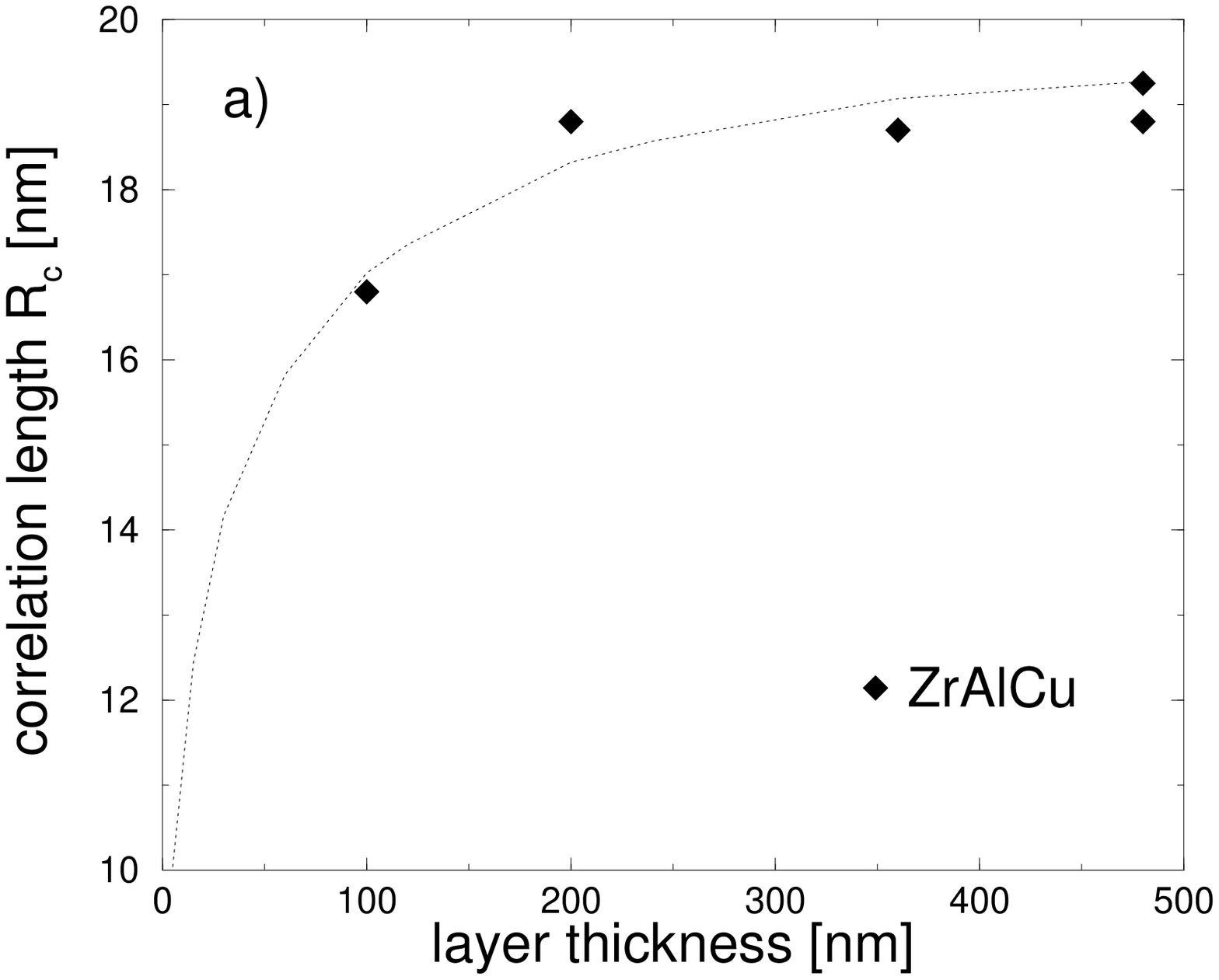}
\hspace*{\fill}
\epsfxsize 7.0cm
\epsfbox{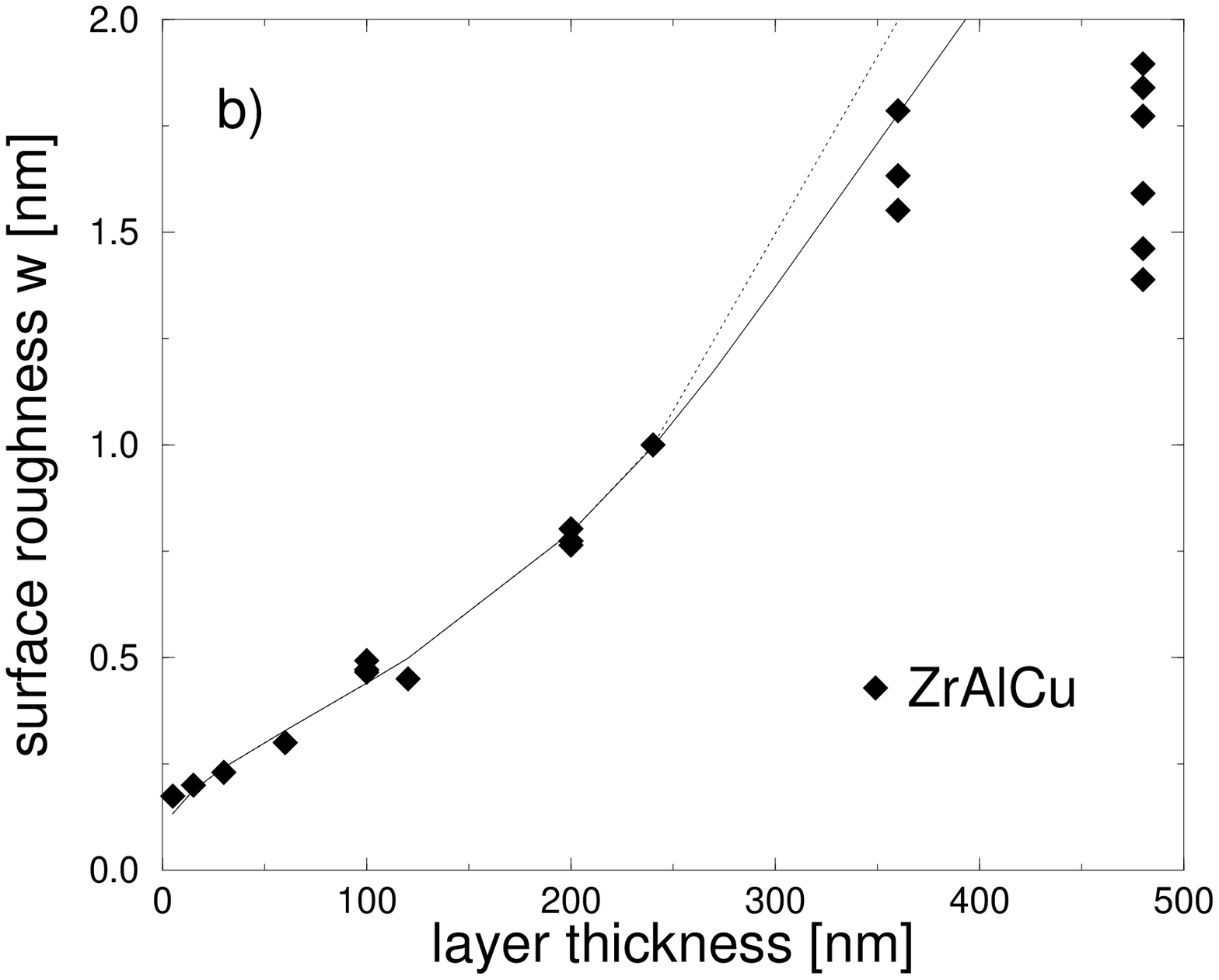}
\hspace*{\fill}
\end{center}
\caption{The correlation length $R_c$ and surface roughness $w$ (dotted lines),
defined below eq. (\ref{corr}), {\em vs.} layer thickness $\left<H\right>$,
see below eq. (\ref{noise}), are
calculated from the linear growth equation (\ref{lin1}) using the parameters
given in eq. (\ref{para}). In fig.~\ref{fig:bildrcw1}b we depict the
influence of the STM tip with an apex angle of $100^{\circ}$ (solid line).
The diamond symbols represent the corresponding experimental results.}
\label{fig:bildrcw1}
\end{figure}
\begin{figure}[htb]
\begin{center}
\hspace*{\fill}
\epsfxsize 7.0cm
\epsfbox{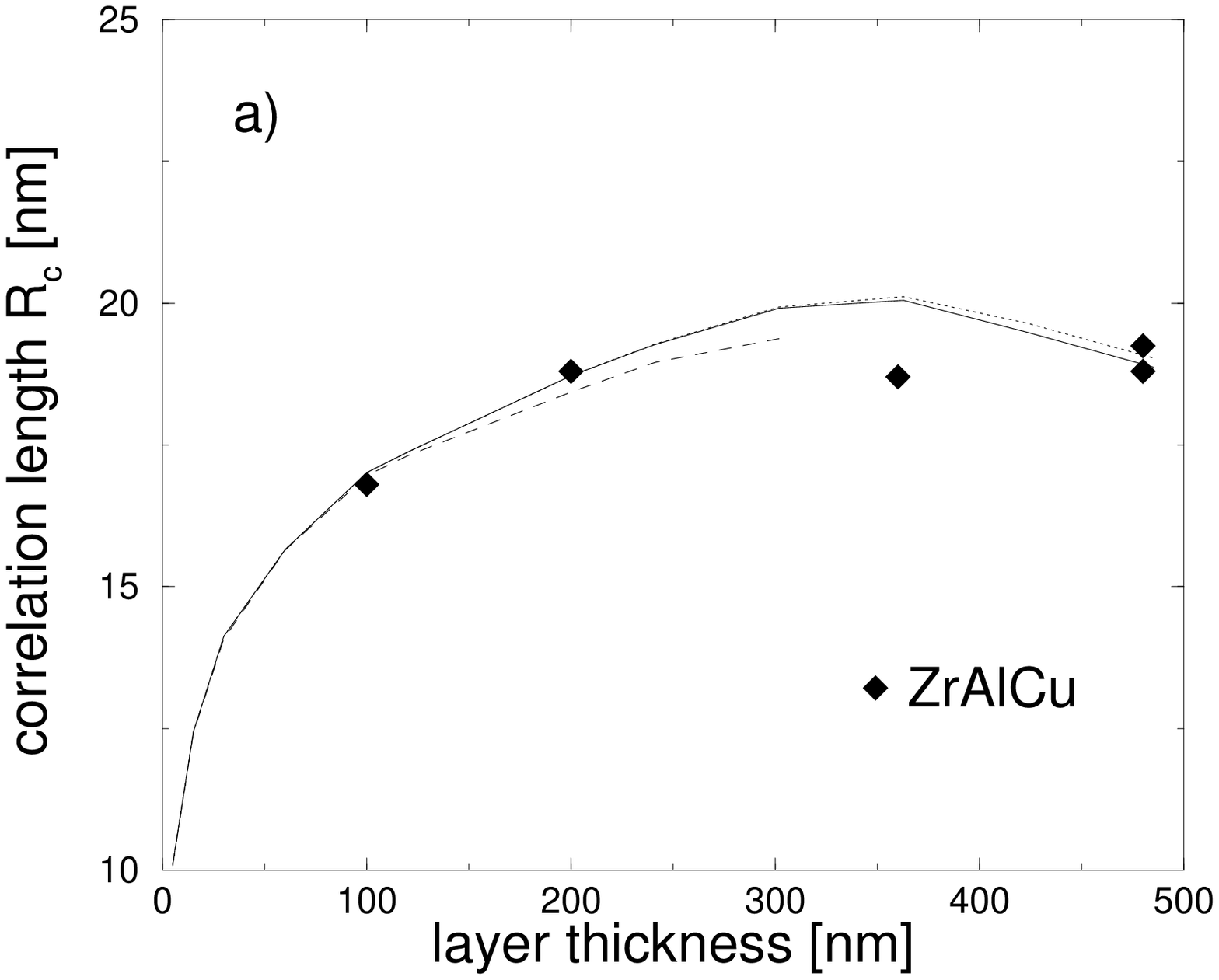}
\hspace*{\fill}
\epsfxsize 7.0cm
\epsfbox{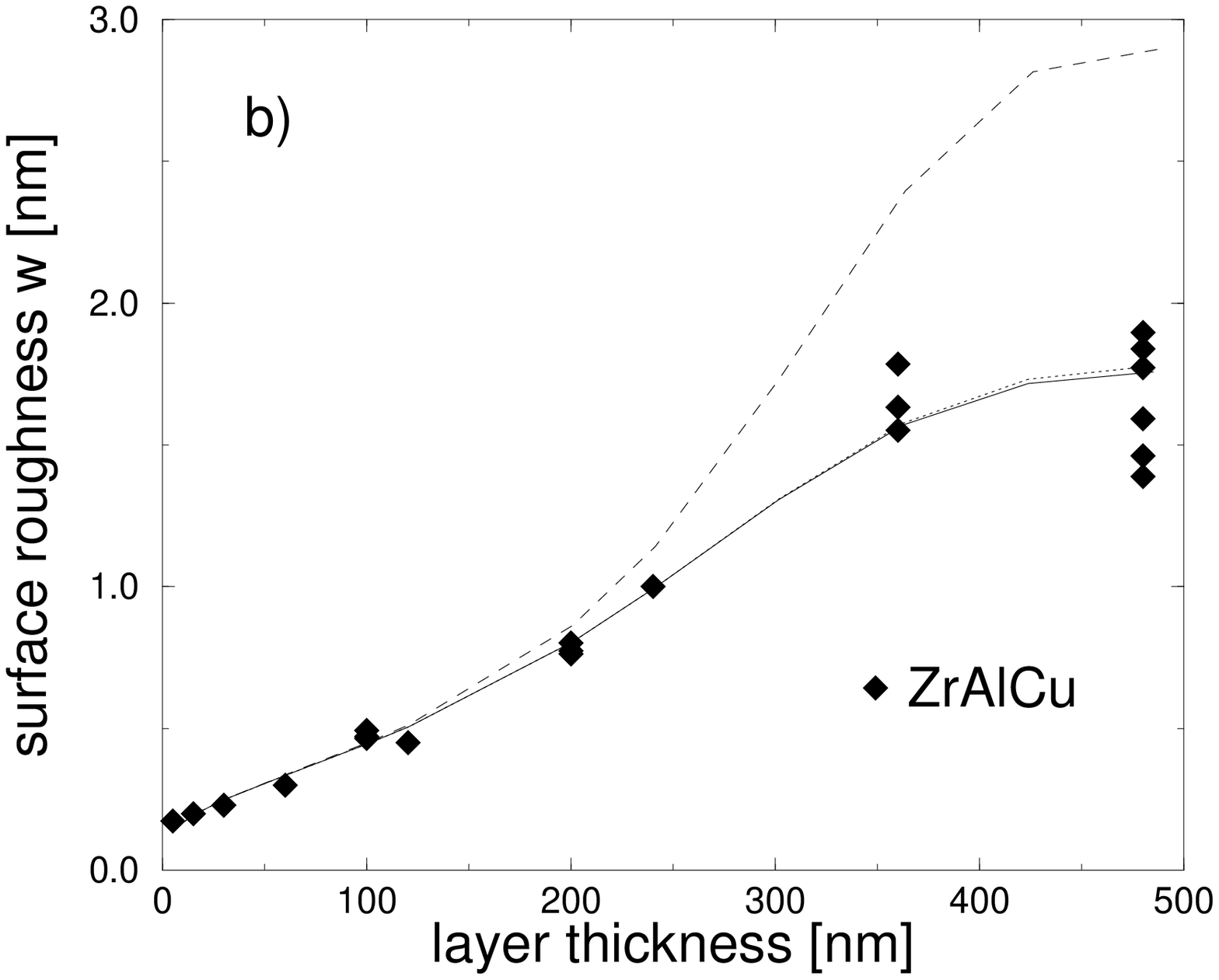}
\hspace*{\fill}
\end{center}
\caption{Correlation length $R_c$ and surface roughness $w$ calculated from
the nonlinear growth equation (\ref{sto2}) using the parameters
$a_3=-0.10nm^3/s$ and $a_4=0.055nm/s$
with (solid lines) and without (dotted lines) consideration of
the STM tip (with an apex angle of $100^{\circ}$).
The parameters of the linear parts of
the equation, $a_1$, $a_2$, and $D$, are given in eq. (\ref{para}).
To demonstrate the physical relevance of the nonlinear growth law
$\propto\nabla^2(\nabla h)^2$ we depict for comparison the prediction that
results by setting $a_3=0$ (dashed lines without consideration of the STM tip).
The diamond symbols represent the corresponding experimental results.}
\label{fig:bildrcw2}
\end{figure}
\end{document}